\documentclass[aps,twocolumn,pra,superscriptaddress,showpacs,showkeys,floatfix]{revtex4-1}
\usepackage{amsbsy} 
\usepackage{graphicx}
\usepackage{amsmath}
\usepackage{amssymb}
\input{epsf}

\begin{document}
\title{Non-perturbative interpretation of the Bloch vector's path 
\\beyond rotating wave approximation}

\author{Giuliano Benenti}
\affiliation{CNISM and Center for Nonlinear and Complex Systems,
Universit\`a degli Studi dell'Insubria, via Valleggio 11, 22100 Como, Italy}
\affiliation{Istituto Nazionale di Fisica Nucleare, Sezione di Milano,
via Celoria 16, 20133 Milano, Italy}
\author{Stefano Siccardi}
\affiliation{Department of Information Technologies, University of Milan,
via Bramante 65, 26013 Crema, Italy}
\author{Giuliano Strini}
\affiliation{Department of Physics, University of Milan,
via Celoria 16, 20133 Milano, Italy}

\begin{abstract}
The Bloch vector's path of a two-level system exposed to a 
monochromatic field exhibits, in the regime of strong 
coupling, complex corkscrew trajectories. By considering
the infinitesimal evolution of the two-level system
when the field is treated as a classical object,
we show that the Bloch vector's rotation speed oscillates
between zero and twice the rotation speed predicted
by the rotating wave approximation. Cusps appear 
when the rotation speed vanishes. We prove analytically 
that in correspondence to cusps the curvature of the
Bloch vector's path diverges. On the other hand, numerical
data show that the curvature is very large even for a
quantum field in the deep quantum regime with mean number
of photons $\bar{n}\lesssim 1$. We finally compute 
numerically the typical error size in a quantum gate 
when the terms beyond rotating wave approximation 
are neglected.
\end{abstract}


\maketitle

\section{Introduction}

The dynamics of two-level systems in an external field has
been thoroughly studied for years. This problem appears in
several physical systems and is significant for the realization
of quantum gates~\cite{qcbook,nielsen}.
The model consists of the Hamiltonian of a
two-level system, coupled to a classical field or to a
quantum harmonic oscillator.
There are no general methods for solving analytically
these models without any approximations, so several
approaches have been used to study their behavior.
The best known is the Rotating Wave 
Approximation (RWA),
that consists in neglecting the effects of the rapidly
rotating terms~\cite{micromaser}.

For an atom residing in a resonant cavity the frequency
$\Omega$ of the Rabi oscillations between the two relevant 
states of the atom is typically $10^{-6}$ of the atomic
frequency $\omega_a$ and of the cavity frequency $\omega$,
so that the RWA yields a good description of the 
system~\cite{haroche}. On the other hand, in circuit 
quantum electrodynamics (cQED)~\cite{blais,wallraff}, 
where superconducting qubits
play the role of artificial atoms, one can enter the 
so-called ultrastrong coupling regime in which the 
ratio $\Omega/\omega>0.1$ 
\cite{bourassa,gross,mooij} (at resonance, $\omega_a=\omega$).
In this regime, effects beyond RWA should be taken into 
account, both for the dynamics of pure 
states~\cite{shahriar,irish,casanova,nori,grifoni,grifoni2}
and for dissipative dynamics~\cite{grifoni3,beaudoin,bina}. 
The RWA is questioned in several contexts, 
including optical forces on two-level atoms~\cite{kumar},
non-Markovian dynamics~\cite{zeng,makela},
entanglement generation~\cite{jing,chen,eberly},
quantum Zeno and anti-Zeno effects~\cite{zheng,ai,cao}
geometric properties of the state evolution~\cite{larson}, and
holonomic quantum computation~\cite{spiegelberg}.

The consideration of high Rabi frequencies is natural in
quantum information theory, since high speed operations are 
needed to perform a large number of quantum gates within the
decoherence time scale, an unavoidable request for 
fault tolerant quantum computation~\cite{qcbook,nielsen}. 
Therefore, a deeper understanding of the effects beyond
RWA is relevant for the prospects of quantum computation.

In this paper, we explain, in a non-perturbative manner, 
the non-trivial Bloch vector's path of a two-level system exposed 
to a monochromatic field.
By considering the infinitesimal evolution of the state 
vector, we determine the temporal dependence of the
rotation axis and frequency. In particular, we show 
analytically that, when the field is treated as a classical
object, the rotation frequency oscillates between zero and 
twice the value predicted by the RWA. The vanishing of
the rotation frequency is associated with cusps in the 
Bloch vector's path. Correspondingly, the curvature 
of the Bloch vector's trajectory diverges.
On the other hand, we show numerically that the 
curvature can take very large values even for a quantum field 
in the deep quantum regime with mean number of photons
$\bar{n}\lesssim 1$. We finally show numerically 
that the size of the errors induced in a quantum gate by the 
terms beyond RWA scales as $1/\omega$.

The paper is organized as follows. The model for a two-level 
system in a classical field is introduced and studied numerically
in Sec.~\ref{sec:classical_model}, while the analytical 
interpretation of the Bloch vector's paths is reported in 
Sec.~\ref{sec:interpretation}. 
Numerical results for the quantum field model
are shown in Sec.~\ref{sec:quantum_model},
where the errors introduced by the terms beyond RWA are
also investigated. 
We finish with concluding remarks in 
Sec.~\ref{sec:conclusions}.

\section{The classical field model}
\label{sec:classical_model}

We consider the following time-dependent Hamiltonian $H(t)$, describing
the interaction of a two-level system with a classical monochromatic 
field (we set $\hbar=1$):
\begin{equation}
  \begin{array}{c}
{\displaystyle
H(t)=H_0+H_I(t),
}
\\
\\
{\displaystyle
H_0=\epsilon_0 |0\rangle\langle 0| + \epsilon_1 |1\rangle\langle 1|,
}
\\
\\
{\displaystyle
H_I(t)= 2 \Omega\cos(\omega t) (|0\rangle\langle 1|+
|1\rangle\langle 0|),
}
\end{array}
\label{eq:hamc}
\end{equation}
where $\omega$ is the frequency of the field 
and $\Omega$ the (Rabi) frequency of the field-induced 
oscillations between the 
two levels $|0\rangle$ and $|1\rangle$ 
\cite{qcbook_ex}.
The time evolution of the two-level state vector
$|\psi(t)\rangle=C_0(t)|0\rangle+C_1(t)|1\rangle$ is,
in the interaction picture, governed by the equations
\begin{equation}
  \left\{
\begin{array}{l}
{\displaystyle
    i  \dot{C}_0(t)  =
    \Omega \left[
      e^{i(\omega-\omega_a)t} +
      e^{-i(\omega+\omega_a)t}
    \right]\,C_1(t),
}
\\
  \\
{\displaystyle
    i  \dot{C}_1(t)  =
    \Omega \left[
      e^{i(\omega+\omega_a)t} +
      e^{-i(\omega-\omega_a)t}
    \right]\,C_0(t),
}
  \end{array}\right.
  \label{eq:noRWA1}
\end{equation}
where 
$\omega_a=\epsilon_1-\epsilon_0$ ($\epsilon_1>\epsilon_0$).

The terms depending on $\omega+\omega_a$ oscillate very rapidly and are
neglected in the rotating wave approximation. 
In this paper, we will explore the effects of these terms
beyond RWA on the evolution of the two-level state vector
$|\psi(t)\rangle$.  
For the sake of simplicity, 
we set the detuning $\Delta=\omega-\omega_a=0$, 
as the treatment for $\Delta\ne 0$ would be essentially identical 
to that for $\Delta=0$. 
Finally, it is convenient to normalize time in units of the 
Rabi frequency,
that is, we set $\Omega=1$.
Hence, we obtain
\begin{equation}
\left\{
\begin{array}{l}
{\displaystyle
i  \, {\dot C}_{0} \;=\; 
 \left( 1 \;+\;  e^{- 2 \,i\, \omega \, t} \right) \, C_{1},
}
\\
\\
{\displaystyle
i  \, {\dot C}_{1} \;=\; 
 \left( 1 \;+\;  e^{2 \,i\, \omega \, t} \right) \, C_{0}.
}
\end{array}
\right.
\label{eq:noRWA}
\end{equation}
The RWA approximation is valid
when $\omega\gg \Omega$
(in our units, when $\omega\gg 1$).

A convenient geometric picture of the evolution of the state 
vector is provided by the Bloch ball 
representation~\cite{qcbook,nielsen}, with the 
Bloch coordinates defined as
\begin{equation}
\left\{
\begin{array}{l}
X = 2 C_{0r}C_{1r} + 2 C_{0i}C_{1i},
\\
\\
Y = 2 C_{0r}C_{1i} - 2 C_{0i}C_{1r},
\\
\\
Z = 2 C_{0r}^{2} + 2 C_{0i}^{2} - 1,
\end{array}
\right.
\label{eq:Bloch}
\end{equation}
where $C_{k r}$ and $C_{k i}$ denote the real and imaginary parts
of $C_k$ ($k=0,1$). The normalization constraint $|C_0|^2+|C_1|^2=1$
implies that the motion of the Bloch vector ${\bf R}=(X,Y,Z)$ 
takes place on the unit (Bloch) sphere $X^2+Y^2+Z^2=1$.

\begin{figure*}[ht]
\includegraphics[angle=0.0, width=16cm]{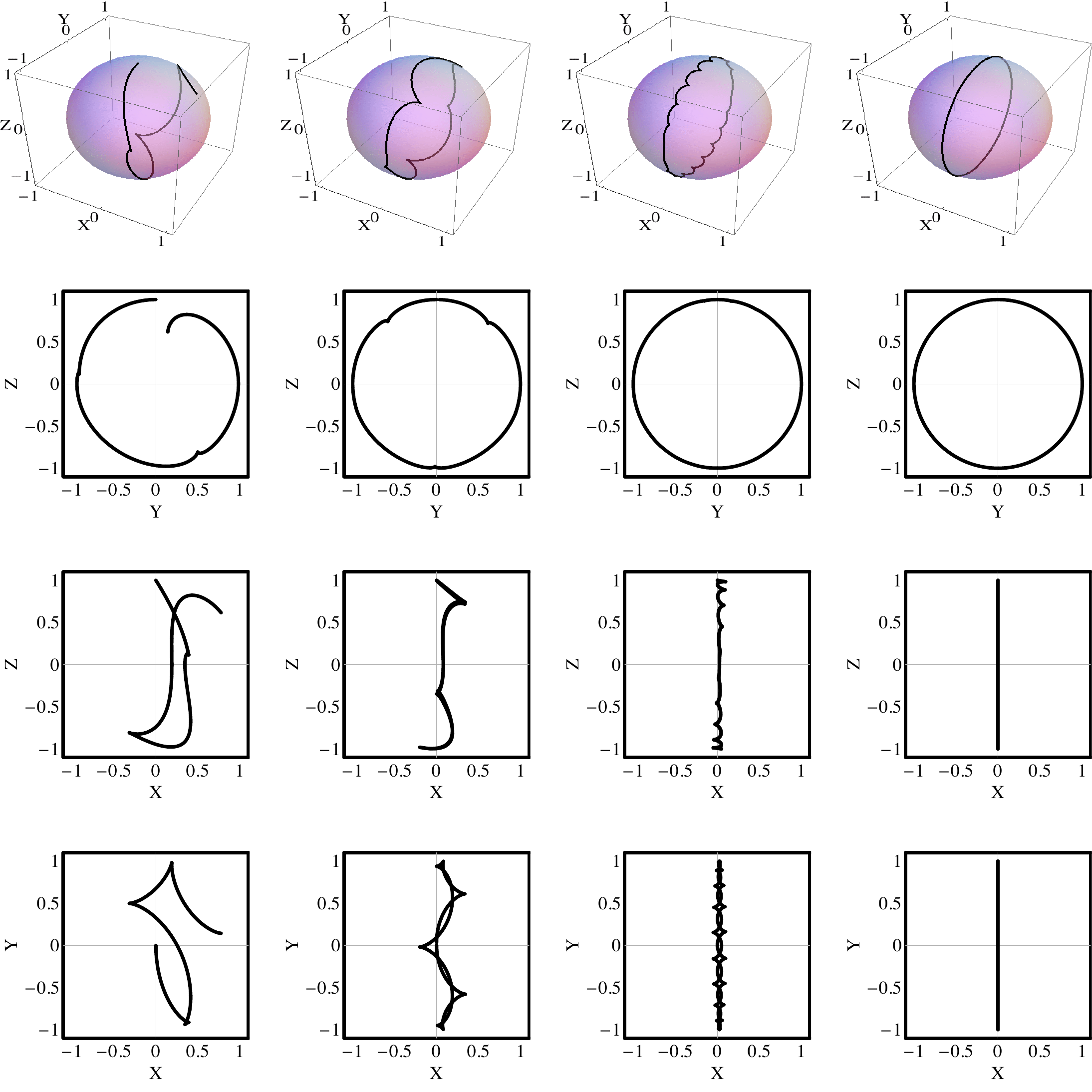}
\caption{Bloch vector's path (top plots) and its projections on,
from the second to the fourth row, the $YZ$, $XZ$, and 
$XY$ planes. From left to right: 
$\omega=2.5,5,20$, and
RWA ($\omega\to\infty$). Time evolution is followed, starting 
from the north pole, up to time $\pi$ (within RWA, 
at such time the Bloch vector returns, for the first time,
back to the north pole). The field is treated as a 
classical object. 
}
\label{fig:fig1}
\end{figure*}

The most interesting feature the Bloch
vector's path is the presence of cusps, shown in Fig.~\ref{fig:fig1}. 
The corkscrew trajectories shown in this figure are obtained 
from numerical integration of the differential 
equations (\ref{eq:noRWA}), starting from the initial condition 
$C_{0}(0)=1$, $C_1(0)=0$ (north pole of the Bloch sphere). 
The plots of Fig.~\ref{fig:fig1} show, from top to bottom, the
Bloch vector's trajectory and its projections on the 
$YZ$, $XZ$, and $XY$ planes; from left to right, 
$\omega=2.5,5,20$, and RWA approximation 
($\omega\to\infty$). 
While within RWA 
the trajectory is a circle, corresponding to 
Rabi oscillations between states $|0\rangle$ and 
$|1\rangle$ (north and south pole of the Bloch sphere,
respectively), cusps appear beyond RWA. 
We will interpret in a simple but exact manner these
numerical results in Sec.~\ref{sec:interpretation}.

\section{Interpretation of the Bloch vector's path}
\label{sec:interpretation}

We consider the infinitesimal evolution of the state vector:
\begin{equation}
|\psi(t+dt)\rangle = U(t,t+dt) |\psi(t)\rangle,
\label{eq:atto_moto1}
\end{equation}
where the unitary operator $U(t,t+dt)$ represents an infinitesimal
rotation of the Bloch vector 
through an angle $d\theta$ about the axis directed along the 
unit vector $\hat{\bf n}=(n_X,n_Y,n_Z)$:
\begin{equation}
U(t,t+dt)=\left(I- i\,\frac{d\theta}{2}\,\hat{\bf n}\cdot 
{\boldsymbol\sigma}\right),
\end{equation}
where $I$ 
is the identity operator and ${\boldsymbol\sigma}=(\sigma_X,\sigma_Y,\sigma_Z)$,
$\sigma_k$ ($k=X,Y,Z$) being the Pauli operators. 
In the $\{|0\rangle,|1\rangle\}$ basis,
Eq.~(\ref{eq:atto_moto1}) reads as follows:
\begin{equation}
\begin{array}{l}
{\displaystyle
\left[
\begin{array}{c}
C_0(t+dt)
\\
\\
C_1(t+dt)
\end{array}
\right]}
\\
\\
{\displaystyle
=
\left[
\begin{array}{cc}
1-i n_Z\,\frac{d\theta}{2} & 
-(n_Y+in_X)\frac{d\theta}{2}
\\
\\
(n_Y-in_X)\frac{d\theta}{2} & 
1+i n_Z\,\frac{d\theta}{2}  
\end{array}
\right]
\left[
\begin{array}{c}
C_0(t),
\\
\\
C_1(t)
\end{array}
\right].
}
\end{array}
\label{eq:atto_moto2}
\end{equation}
Since
\begin{equation}
|\psi(t+dt)\rangle=|\psi(t)\rangle + |\dot{\psi}(t)\rangle\,dt,
\label{eq:atto_moto3}
\end{equation}
with $|\dot{\psi}(t)\rangle$ obtained from Eq.~(\ref{eq:noRWA}), we have
\begin{equation}
\begin{array}{l}
{\displaystyle
\left[
\begin{array}{c}
C_0(t+dt)
\\
\\
C_1(t+dt)
\end{array}
\right]
}
\\
\\
{\displaystyle
=
\left[
\begin{array}{cc}
1 & -i\left(1+e^{-2i\omega t}\right)
\\
\\
-i\left(1+e^{2i\omega t}\right) & 1
\end{array}
\right]
\left[
\begin{array}{c}
C_0(t),
\\
\\
C_1(t)
\end{array}
\right].
}
\end{array}
\label{eq:atto_moto4}
\end{equation}

From comparison between Eqs.~(\ref{eq:atto_moto2}) and (\ref{eq:atto_moto4})
we obtain
\begin{equation}
\left\{
\begin{array}{l}
{\displaystyle
n_{X} \frac{d \theta}{2} \,=\, [ 1+ \cos( 2 \omega t) ] dt, 
}
\\  
\\
{\displaystyle
n_{Y} \frac{d \theta}{2} \,=\, \sin( 2 \omega t) dt,
}
 \\
 \\
{\displaystyle
n_{Z} \,=\, 0.
}
\end{array}
\right.
\label{eq:nxyz}
\end{equation}
From the first two equations of this system we have 
\begin{equation}
\dot{\theta}=\frac{d\theta}{dt} = 4 |\cos(\omega t)|.
\label{eq:dthetadt}
\end{equation}
Therefore, the rotation speed $\dot{\theta}$ vanishes when 
$\cos(\omega t)=0$, that is, for 
\begin{equation}
t=t_k=\frac{(2k+1)\,\pi}{2\,\omega}, \quad (k=0,1,...).
\label{eq:tk}
\end{equation}
After insertion of Eq.~(\ref{eq:dthetadt}) into
Eq.~(\ref{eq:nxyz}), we obtain
\begin{equation}
\left\{
\begin{array}{l}
{\displaystyle
n_{X} \,=\, |\cos(\omega t)|,
}
\\  
\\
{\displaystyle
n_{Y}  \,=\, \sin(\omega t)\,{\rm sgn}\,[\cos(\omega t)],
}
\\  
\\
{\displaystyle
n_{Z} \,=\, 0,
}
\end{array}
\right.
\label{eq:nvec}
\end{equation}
where ${\rm sgn}(x)=x/|x|$ is the sign function.
Therefore, at times $t_k$ the rotation axis changes instantaneously 
from $(n_X,n_Y,n_Z)=(0,1,0)$ to $(0,-1,0)$, that is, from 
the $Y$ axis to the opposite direction $-Y$. 
Such discontinuity in the rotation 
axis is associated with a cusp in the Bloch vector's path. 
As discussed in appendix~\ref{sec:curvature}, the curvature
of the Bloch vector's path diverges at the cusps.  

\begin{figure}[ht]
\includegraphics[angle=0.0, width=8cm]{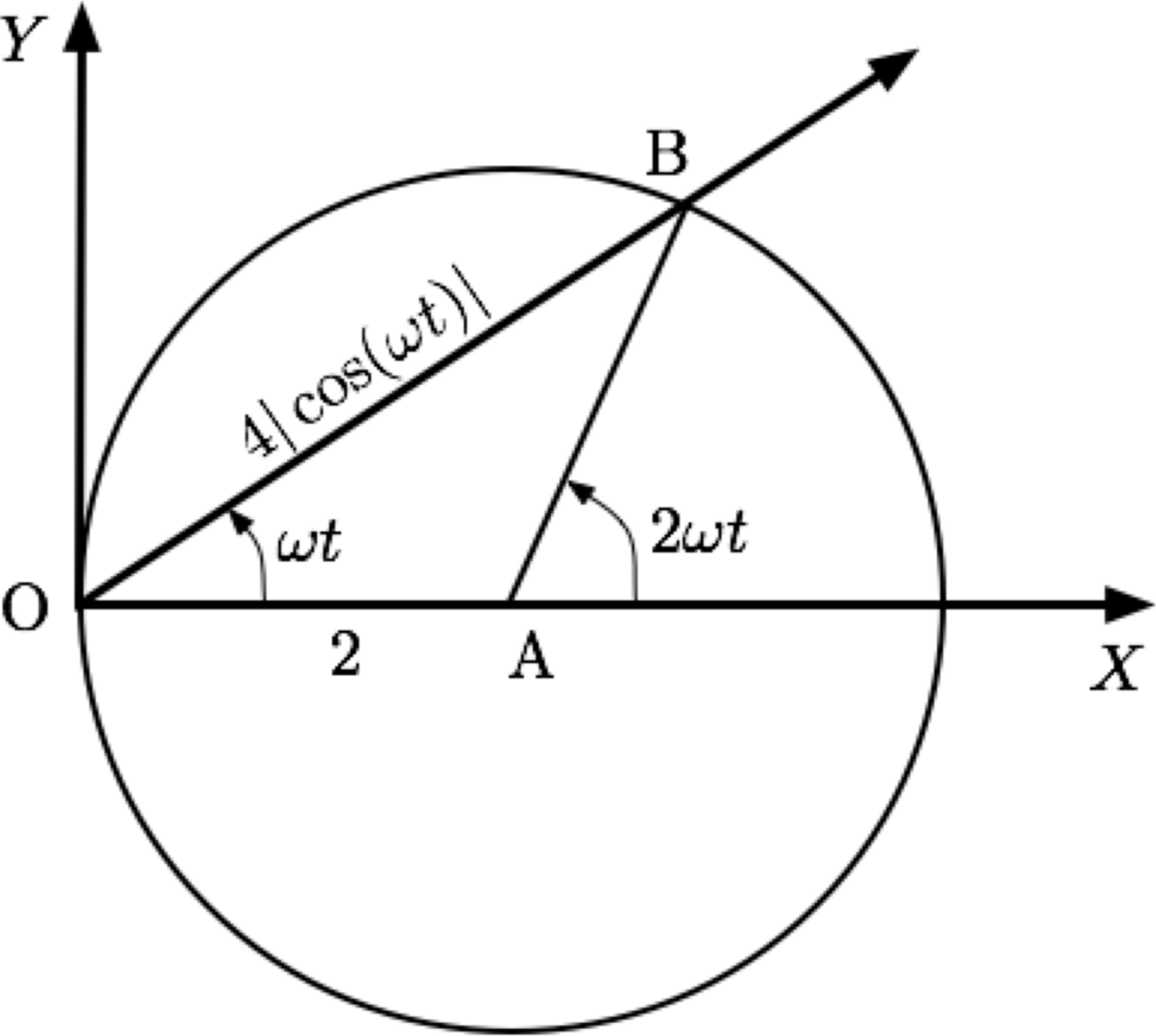}
\caption{Schematic illustration of the rotation 
axis and speed within and beyond RWA.}
\label{fig:fig2}
\end{figure}

The dependence of the rotation axis and speed on time can 
be visualized by means of the schematic drawing in 
Fig.~\ref{fig:fig2}. Within RWA, the rotation speed is 
given by the length $\dot{\theta}_{\rm RWA}=2\Omega=2$ 
($\Omega=1$ in our units) of the segment
$\hbox{OA}$ and the rotation axis is the $X$ axis. 
On the other hand, when effects beyond RWA are
taken into account, the rotation speed at time $t$ is 
given by the length $\dot{\theta}=4|\cos(\omega t)|$ of the
segment $\hbox{OB}$, and the rotation axis is directed along
$\hbox{OB}$. Therefore, the rotation speed oscillates between
$\dot{\theta}=0$ (at times $t_k$) and 
$\dot{\theta}=4=2\dot{\theta}_{\rm RWA}$
(at times $\tilde{t}_k=k \pi/\omega$, $k=0,1,...$). 

\begin{figure}[ht]
\includegraphics[angle=0.0, width=8cm]{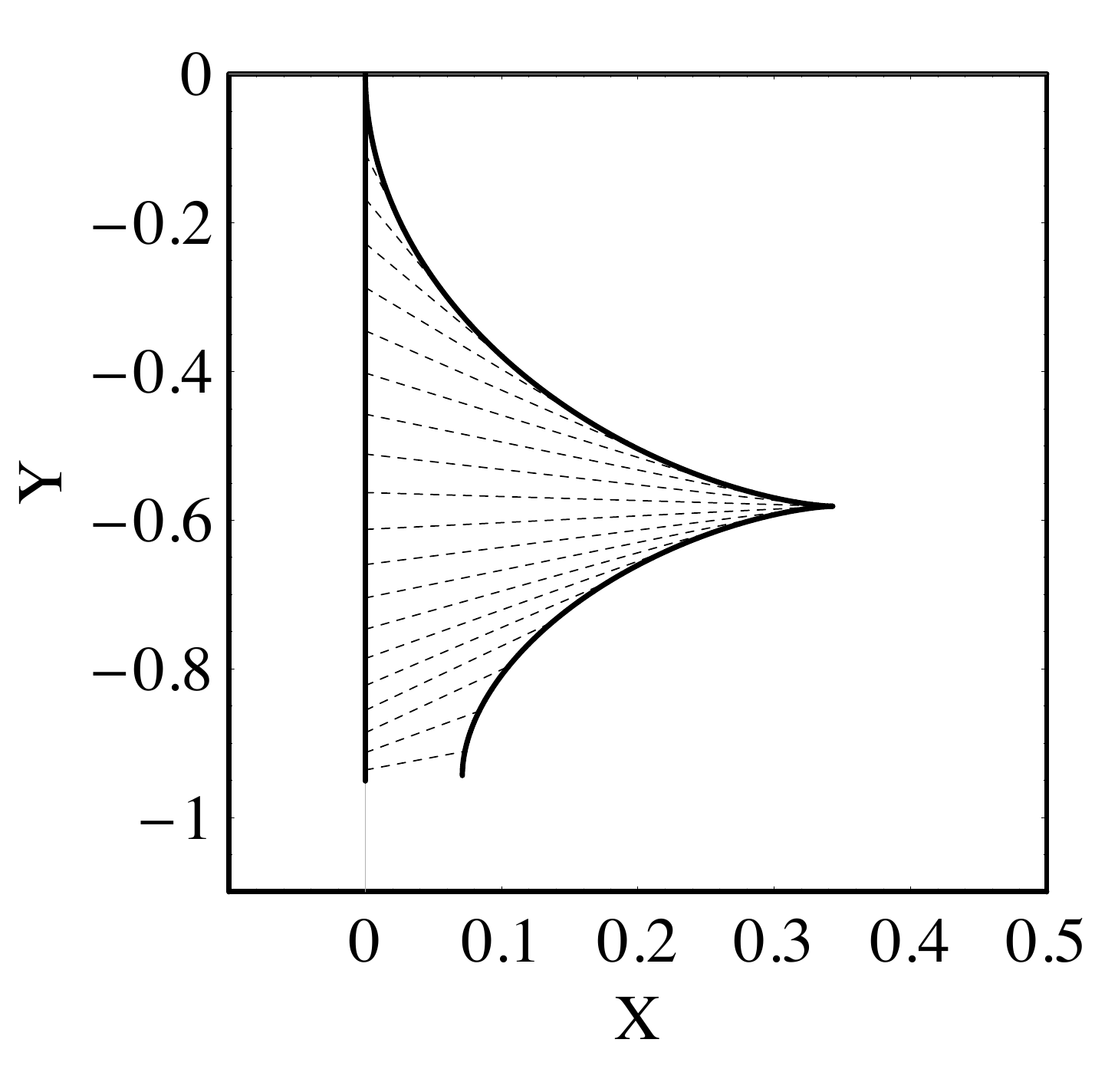}
\caption{Comparison between the 
$XY$ projection of the Bloch vector's 
exact evolution
(full curve) and its RWA (straight line) in the 
vicinity of a cusp, for $\omega=5$. 
The dashed lines link points corresponding to 
the exact and approximate Bloch vectors, 
${\bf R}(t)$ and ${\bf R}_{\rm RWA}(t)$, obtained 
after the same evolution time $t$ (the different dashed lines
correspond to different values of $t$).} 
\label{fig:fig3}
\end{figure}

In Fig.~\ref{fig:fig3}, we show 
the $XY$-plane projection of the Bloch vector's evolution in
the neighborhood of a cusp and compare it with the RWA approximation.
It is interesting to remark that the distance between the two 
evolutions, shown at different times by dashed lines, 
is not simply given by the shortest distance between
the exact trajectory and its RWA. This is due to the
fact that, as pointed out above, 
while the RWA predicts a constant rotation 
velocity $\dot{\theta}_{\rm RWA}=2$, 
the exact evolution exhibits a variable rotation speed 
$\dot{\theta}=4|\cos(\omega t)|$, which is faster than 
in the RWA far from the cusp and slows down up to $\dot{\theta}=0$
at the cusps. 

\begin{figure}[ht]
\includegraphics[angle=0.0, width=8cm]{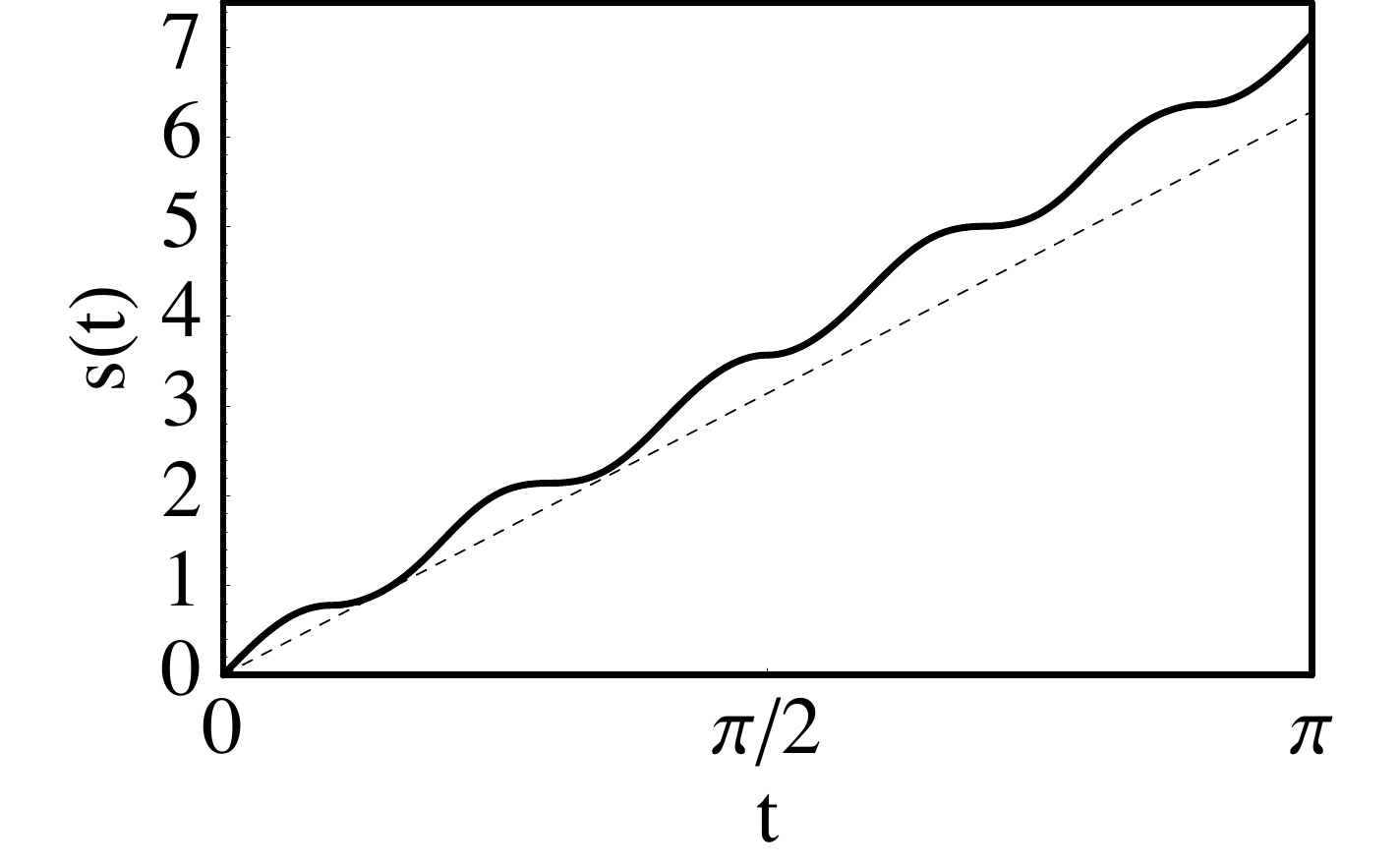}
\caption{Plateaus in the path length $s(t)$ as a function of 
time (full curve), for $\omega=5$. 
The straight dashed line corresponds to the 
RWA result, $s(t)=2 t$.}
\label{fig:fig4}
\end{figure}

The above discussed non-trivial behavior of the Bloch vector's path 
implies, as shown in Fig.~\ref{fig:fig4}, 
the existence of plateaus in the dependence of the path 
length $s(t)$ as a function of time. The plateaus are
obtained around the times $t_k$ where the curvature
of the Bloch vector's path diverges. The slope of 
$s(t)$ oscillates between zero (at the cusps) and 
$2\dot{\theta}_{\rm RWA}=4$. Note that the overall path length
is larger than within RWA, since the Bloch vector moves
along a path longer than within RWA (see Fig.~\ref{fig:fig3}).
The mean slope of $s(t)$ approaches the 
RWA slope in the limit $\omega\to\infty$.

Finally, we point out that, while the curvature
exhibits a singularity at the cusps, the functions 
$X(t),Y(t)$, and $Z(t)$ are regular at the times $t=t_k$
corresponding to such cusps. This is due to the fact that,
as discussed above, the rotation speed tends to zero 
when $t\to t_k$~\cite{cauchypeano}.  

\section{The quantum field model}
\label{sec:quantum_model}

The interaction between a two-level system and a single mode of
the quantized electromagnetic field is described by the Hamiltonian
($\hbar=1$)~\cite{micromaser}
\begin{equation}
  \begin{array}{c}
{\displaystyle
H=H_0+H_I,
}
\\
\\
{\displaystyle
H_0=\frac{1}{2}\,\omega_a \sigma_Z + 
\omega\left(a^\dagger a +\frac{1}{2}\right),
}
\\
\\
{\displaystyle
H_I=\lambda \sigma_+\,(a^\dagger+a),
+\lambda^\star \sigma_-\,(a^\dagger+a),
}
\end{array}
\label{eq:noREWAquantum}
\end{equation}
where $\sigma_\pm = \frac{1}{2}\,(\sigma_X\pm i \sigma_Y)$
are the rising and lowering operators for the two-level system:
$\sigma_+ |0\rangle = |1\rangle$,
$\sigma_+ |1\rangle = 0$,
$\sigma_- |0\rangle = 0$,
$\sigma_- |1\rangle = |0\rangle$,
the operators $a^\dagger$ and $a$ create 
and annihilate a photon: 
$a^\dagger |n\rangle_p=\sqrt{n+1}|n+1\rangle_p$, 
$a |n\rangle_p=\sqrt{n}|n-1\rangle_p$, 
$|n\rangle_p$ being the Fock state with $n$ photons. 
For simplicity's sake, we
consider the resonant case ($\omega=\omega_a$) and
the coupling constant $\lambda\in\mathbb{R}$. 
The RWA is obtained when we neglect the term 
$\sigma_+ a^\dagger$, which simultaneously 
excites the two-level system and creates a photon,
and $\sigma_- a$, which de-excites the 
two-level system and 
annihilates a photon. In this limit, Hamiltonian
(\ref{eq:noREWAquantum}) reduces to the Jaynes-Cummings 
Hamiltonian (see, for instance, Ref.~\cite{micromaser}).
Within RWA (Jaynes-Cummings limit), 
there are coherent Rabi oscillations between the 
two-level system-cavity
states $|0\rangle |n+1\rangle_p$ and 
$|1\rangle |n-1\rangle_p$. The frequency of 
such oscillations is $\Omega_n=\lambda \sqrt{n}$.

We study numerically the temporal evolution of the 
two-level system's state vector, when the initial state is 
$|\phi_0\rangle|\alpha\rangle_p$,
where
$|\phi_0\rangle=\cos\left(\frac{\theta}{2}\right)|0\rangle+
\sin\left(\frac{\theta}{2}\right)|1\rangle$
is the initial state of the two-level system
and $|\alpha\rangle_p=\sum_{n=0}^\infty c_n |n\rangle_p$,
with $c_n=\exp(-\frac{1}{2}\,|\alpha|^2)
\frac{\alpha^n}{\sqrt{n !}}$, is the
coherent state of the field,
with mean number of photons given by $\bar{n}=|\alpha|^2$ and
root mean square deviation in the photon number 
$\Delta n=\sqrt{\bar{n}}$. The classical field limit is 
obtained for $\bar{n}\to\infty$, with 
constant $\Omega_{\bar{n}}=\lambda\sqrt{\bar{n}}=\lambda\alpha$, 
that is, with $\lambda\propto 1/\alpha$. We measure
frequencies in units of $\Omega_{\bar{n}}$, namely we
set $\Omega_{\bar{n}}=1$, i.e., $\lambda=1/\alpha$.  

\begin{figure*}[ht]
\includegraphics[angle=0.0, width=16cm]{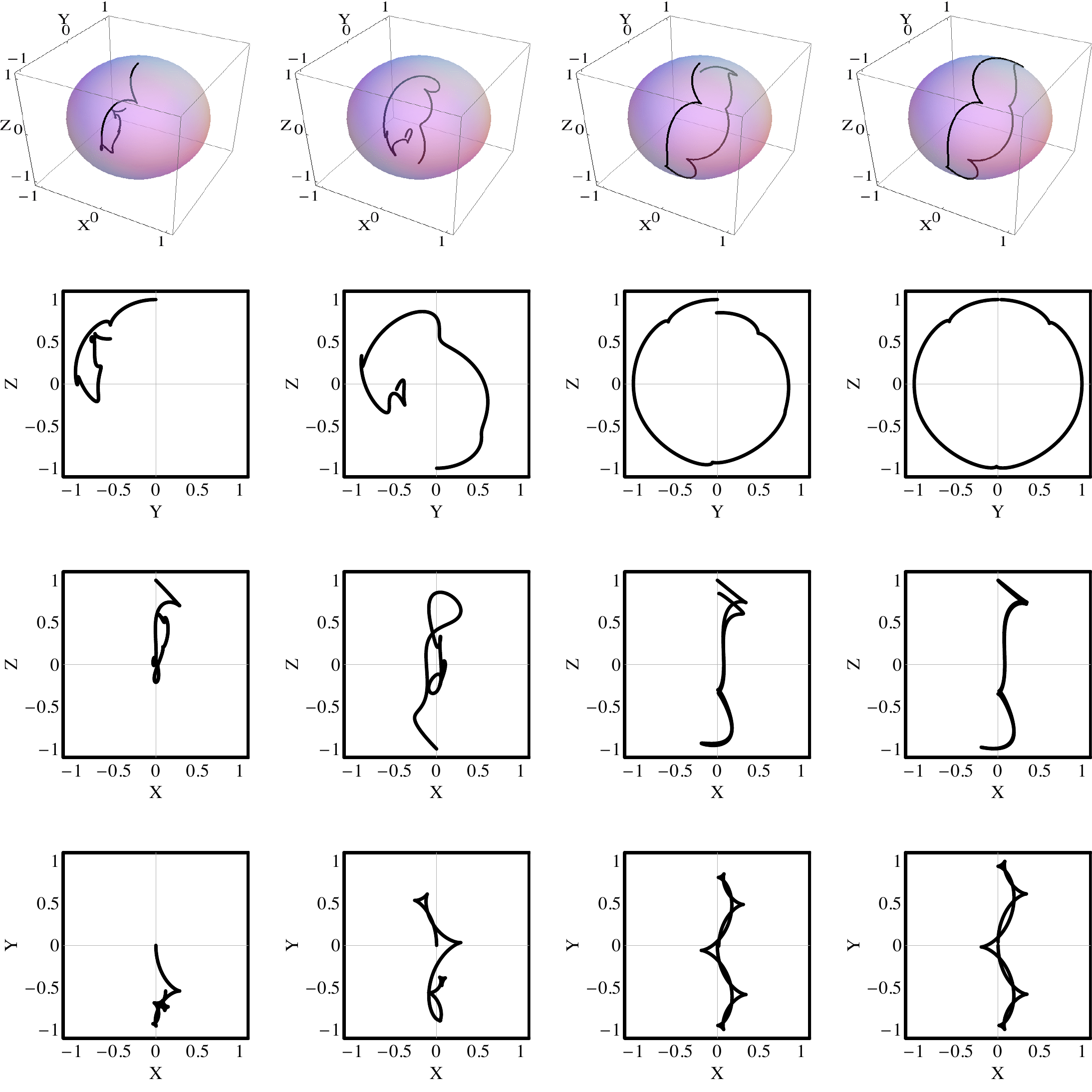}
\caption{Bloch vector's path (top plots) and its projections on,
from the second to the fourth row, the $YZ$, $XZ$, and 
$XY$ planes. From left to right: coherent quantum field with 
$\alpha=1,\theta=0$; 
$\alpha=1,\theta=\pi$; 
$\alpha=5,\theta=0$; and 
classical field with $\theta=0$.
In all instances, $\omega=5$ and 
time evolution is followed up to time $\pi$.}
\label{fig:fig5}
\end{figure*}

We focus on the time evolution of the Bloch 
ball coordinates. In Fig.~\ref{fig:fig5} we show the Bloch 
vector's trajectory and its projection on the 
$YZ$, $XZ$, and $XY$ planes, both in the deep quantum 
regime $\alpha=1$ (first two columns in 
Fig.~\ref{fig:fig5}) and closer to the classical
field limit ($\alpha=5$, corresponding to $\bar{n}=25$ photons,
in the third column of Fig.~\ref{fig:fig5}). 
For small values of $\alpha$ the Bloch vector's path
crucially depends on the initial state $|\phi_0\rangle$ of 
the two-level system, as we can see from the two  
cases shown in Fig.~\ref{fig:fig5}:
$\theta=0$ (north pole of the Bloch sphere) 
and $\theta=\pi$ (south pole) 
(such strong dependence does not appear for large
values of $\alpha$). 
In all instances, $\omega=5$, so that the
effects beyond RWA are pronounced. 
The plots for $\alpha=5$ are quite close to the 
classical field plots (fourth column in 
Fig.~\ref{fig:fig5}). On the other hand, cusps, even though
smoothed, are very pronounced even in the deep quantum regime 
$\alpha=1$. 
In Table~\ref{table:cusps} we show 
the value of the curvature $\kappa$ 
at times $t_1=\frac{\pi}{2\omega}$ and 
$t_2=\frac{3\pi}{2\omega}$, where $\kappa$ diverges
for the classical field model 
(see appendix ~\ref{sec:curvature}). 
Note that the obtained values are much larger than 
the value $\kappa=1$ corresponding to RWA in the classical 
field model. 

\begin{table}[h]
\begin{center}
\begin{tabular}{|c|c|c|}
\hline
  & {\rm first peak}& {\rm second peak}\\
\hline
 $\alpha=1,\theta=0$  & $2.6\times 10^4$ & $1.8\times 10^2$ \\
 $\alpha=1,\theta=\pi$  & $12$ & $8.5$ \\
 $\alpha=5,\theta=0$  & $1.5\times 10^5$ & $2.8\times 10^3$ \\
\hline
\end{tabular}
\end{center}
\caption{Value of the curvature $\kappa$ for
the first two peaks, at times 
$t_1=\frac{\pi}{2\omega}$ and 
$t_2=\frac{3\pi}{2\omega}$ 
where $\kappa$ diverges when the field 
is treated as a classical object.
The three lines of this table correspond to the 
first three columns of Fig.~\ref{fig:fig5}, 
so that $\omega=5$.}
\label{table:cusps}
\end{table}

\begin{figure*}[ht]
\includegraphics[angle=0.0, width=16cm]{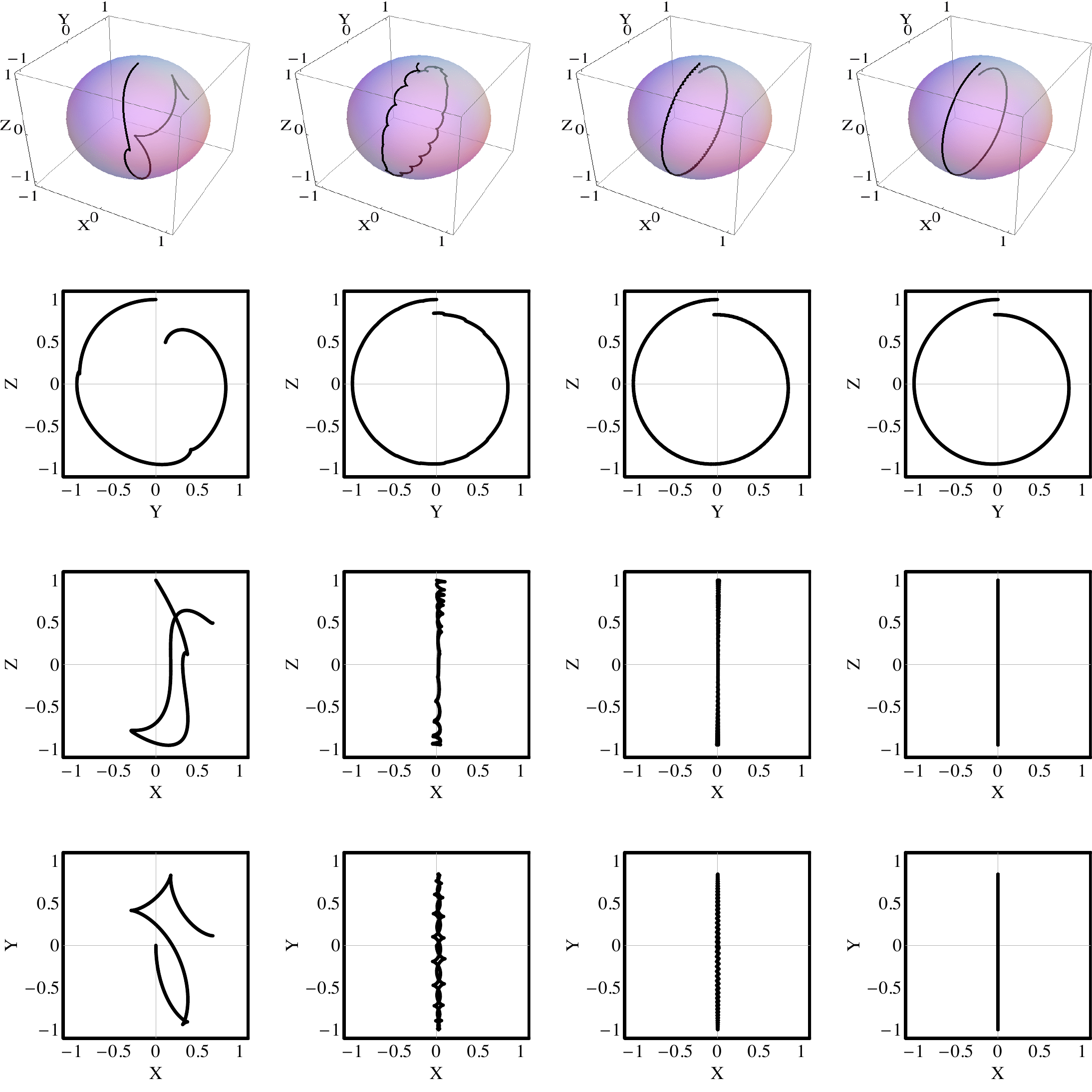}
\caption{Bloch vector's path (top plots) and its projections on,
from the second to the fourth row, the $YZ$, $XZ$, and 
$XY$ planes. From left to right: 
$\omega=2.5$, $20$, $100$ and Jaynes-Cummings Hamiltonian
($\omega= \infty$).
In all instances, we consider a coherent quantum 
field with $\alpha=5,\theta=0$.
Time evolution is followed up to time $\pi$.}
\label{fig_new}
\end{figure*}

We also investigate, for a given value of
the mean number of photons $\bar{n}=|\alpha|^2$, 
the Jaynes-Cummings limit, i.e. $\omega\to\infty$.
In that limit, Hamiltonian
(\ref{eq:noREWAquantum}) reduces to the Jaynes-Cummings
Hamiltonian as the rapidly oscillating (in the interaction 
picture) terms $\sigma_+ a^\dagger$ and $\sigma_- a$ 
can be neglected.  
In Fig.~\ref{fig_new} we show the Bloch
vector's trajectory and its projection on the
$YZ$, $XZ$, and $XY$ planes for a fixed $\bar{n}=25$
and, from left to right, $\omega=2.5$, $20$, $100$, and 
for the Jaynes-Cummings Hamiltonian ($\omega = \infty$). 
Note that data at $\omega=100$ are already quite close
to those obtained for the Jaynes-Cummings Hamiltonian.
As for the classical field within RWA,
in the Jaynes-Cummings limit $\omega\to\infty$ the cusps 
in the Bloch vector's path disappear. 
There is, however, an important difference between the
classical field and the Jaynes-Cummings results. In the 
latter case, the Bloch vector's trajectories are not closed.
This is a consequence of the non-unitary evolution of the 
two-level system exposed to a quantum field, as described by Hamiltonian
(\ref{eq:noREWAquantum}). The Bloch vector's path is 
not bounded to remain on the surface of the Bloch ball 
and exhibits, as well-known in 
the literature~\cite{gea-banacloche,buzek}, a motion similar 
to a spiral and collapses to the center of the Bloch ball,
with repeated revivals (to the surface) and 
collapses at longer times.  
Our data for the Jaynes-Cummings model correspond to the beginning
of the first collapse.

\begin{figure}[ht]
\includegraphics[angle=0.0, width=8cm]{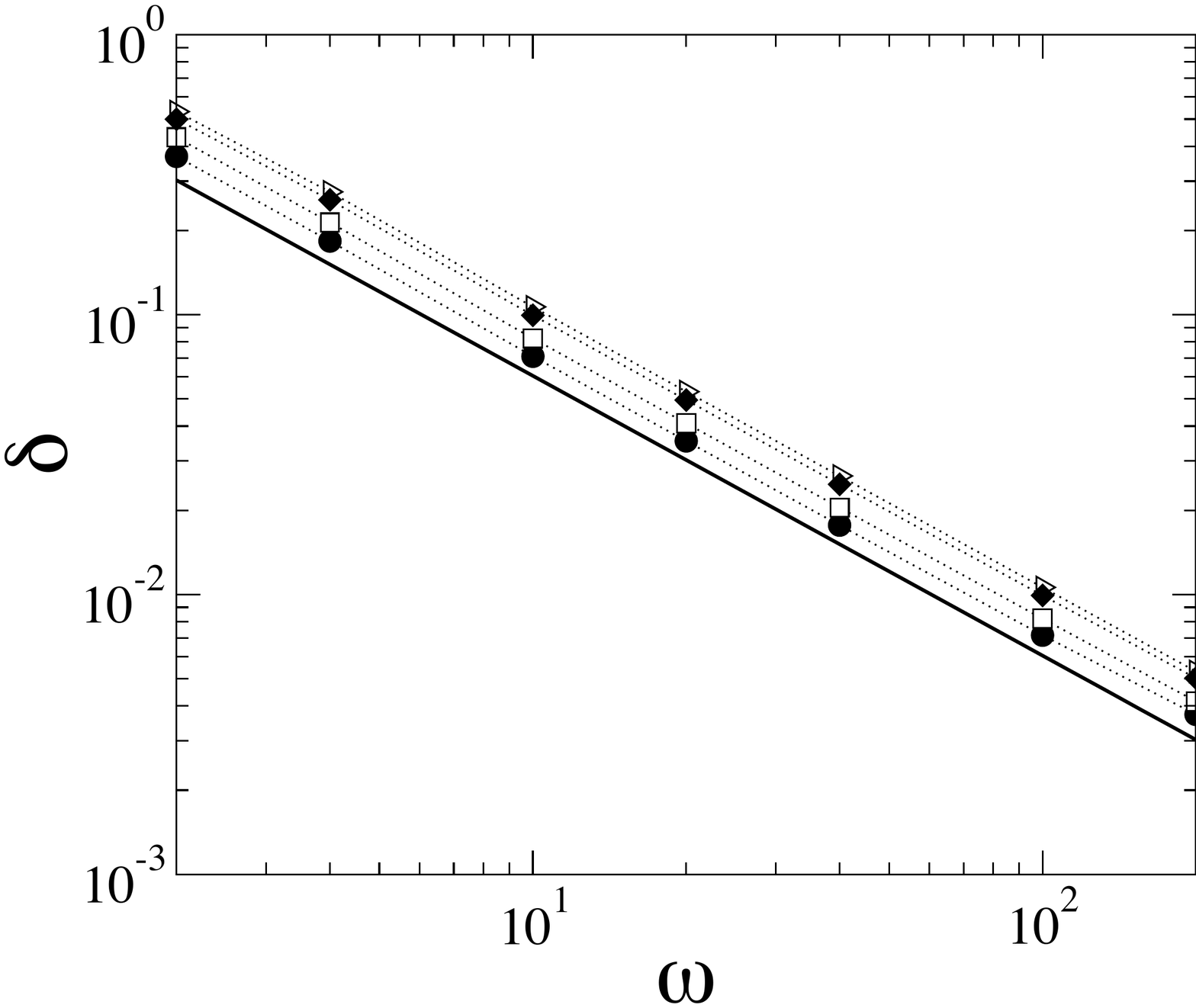}
\caption{Root mean square distance between the exact evolution and
the RWA, for, from bottom to top, 
$\alpha=1$ (circles), 2 (squares), 5 (diamonds) and 
classical field (triangles). The straight full line shows the 
$1/\omega$ slope.}
\label{fig:error}
\end{figure}

We finally estimate the error introduced by the terms beyond
RWA in a typical quantum gate. Since the time elapsed during
an elementary quantum gate is of the order of the inverse of
the Rabi frequency, we compute the Euclidean square distance 
on the Bloch ball between the exact and the RWA evolution,
\begin{equation}
\begin{array}{c}
||{\bf R}-{\bf{R}}_{\rm RWA}||=
\\
\\
\sqrt{(X-X_{\rm RWA})^2+(Y-Y_{\rm RWA})^2+
(Z-Z_{\rm RWA})^2}
\end{array}
\end{equation}
as a function of time, up to $t=\tau=\pi/\Omega$. 
The root mean square 
\begin{equation}
\delta=
\sqrt{\frac{1}{\tau}\int_0^{\tau}
\,dt ||{\bf R}(t)-{\bf{R}}_{\rm RWA}(t)||^2}
\end{equation}
indicates the typical size of the errors 
introduced in a quantum gate by the terms beyond RWA, 
once such terms are neglected.
Our numerical data shown in Fig.~\ref{fig:error} show that
$\delta\propto 1/\omega$ for any value of $\alpha$. 
Moreover, the values obtained for a quantum treatment of 
the field approach quite quickly the expectation of the 
classical field model, and already for $\alpha=5$ we obtain 
values of $\delta$ close to the classical values.

\section{Conclusions}
\label{sec:conclusions}

In this paper, we have explained features 
beyond RWA
of the Bloch vector's path of a two-level system exposed
to a monochromatic field, including the oscillation
of the rotation frequency and axis and the presence of cusps 
in the trajectory. 
Our exact and non-perturbative analysis is based on
the infinitesimal evolution of the two-level system
state vector.

The above non-trivial features of the Bloch vector's path
should be taken into account, together with the effects
of noise, in any implementation of 
quantum gates, where error rates smaller than $10^{-4}$
are requested for fault-tolerant quantum computation. 
In particular, since cusps are clearly seen even for the quantum field
in the deep quantum regime with a small mean number of photons
$\bar{n}\lesssim 1$, they should be relevant and observable in 
circuit QED experiments. 
The numerical results of Fig.~\ref{fig:error} show that 
neglecting the terms beyond RWA one introduces an error of the
order of the inverse of the ratio between the field and the Rabi 
frequency. Therefore, to obtain error rates smaller than 
$10^{-4}$, effects beyond RWA should be necessarily taken into
account when $\Omega/\omega>10^{-4}$, a limit largely 
overtaken in the ultrastrong
in which the ratio $\Omega/\omega>0.1$.
Finally, the existence of cusps due to the presence of two
frequencies, in this case $\omega-\omega_a$ and 
$\omega+\omega_a$, appears as a rather general phenomenon,
so that we expect cusps to appear in protocols such as  
the Raman transition~\cite{footnote}. This remains to 
be analyzed in future~\cite{in_preparation}.

\begin{acknowledgments}
G.B. acknowledges the support by MIUR-PRIN project
``Collective quantum phenomena: From strongly correlated systems to
quantum simulators''.
\end{acknowledgments}

\appendix

\section{Curvature of the Bloch vector's path}
\label{sec:curvature}

Given a point ${\bf R}=(X,Y,Z)$ of the Bloch vector's trajectory, 
we compute its velocity as
\begin{equation}
{\bf V}=\dot{\bf R}={\bf R}\times \dot{\theta}\,\hat{\bf n},
\label{eq:Vvec}
\end{equation}
with $\hat{\bf n}=(n_X,n_Y.n_Z)$ direction of the rotation axis
and $\dot\theta$ rotation frequency.  
After substitution of  Eqs.~(\ref{eq:dthetadt}) and (\ref{eq:nvec}) 
into (\ref{eq:Vvec}), we obtain the components of vector $\vec{V}$:
\begin{equation}
\left\{
\begin{array}{l}
{\displaystyle
V_X=-4Z\sin(\omega t)\cos(\omega t),
}
\\
\\
{\displaystyle
V_Y=4Z\cos^2( \omega t),
}
\\
\\
{\displaystyle
V_Z=4\cos(\omega t)\,[X\sin(\omega t)-Y\cos(\omega t)].
}
\end{array}
\right.
\end{equation}
We then compute the tangent vector 
\begin{equation}
\hat{\bf t} =\frac{d {\bf R}}{ds}=\frac{{\bf V}}{\dot{s}},
\end{equation}
where $s$ is the arc length along the Bloch vector's path and
\begin{equation}
\dot{s}=|{\bf V}|=
4 |\cos(\omega t)| \sqrt{Z^2+[X\sin(\omega t)-Y\cos(\omega t)]^2}.
\end{equation}

\begin{figure}[ht]
\includegraphics[angle=0.0, width=8cm]{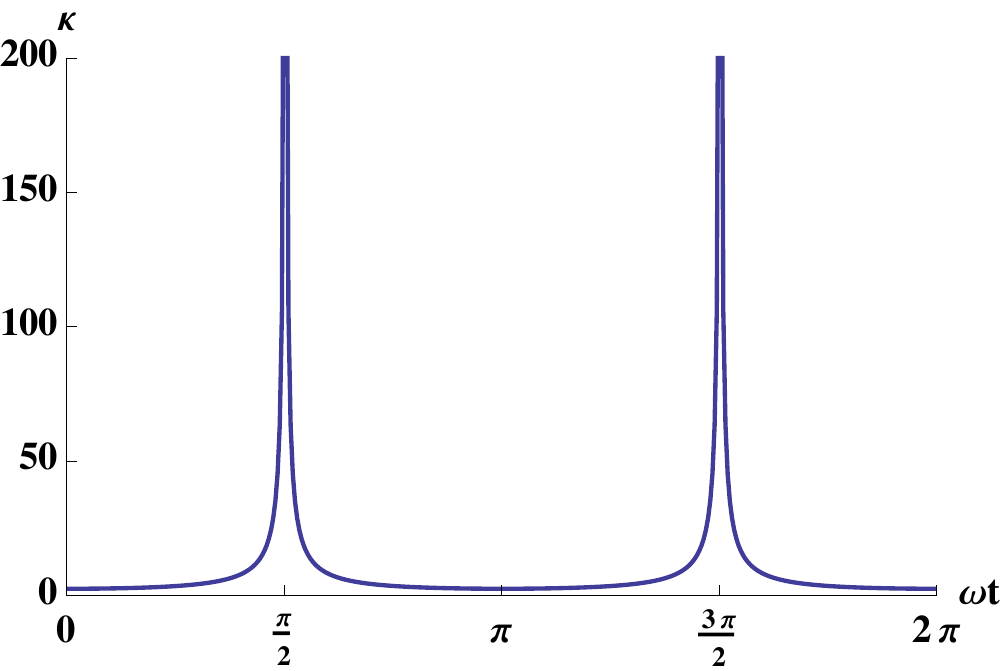}
\caption{Curvature $\kappa$ versus rescaled time
$\omega t$.} 
\label{fig:curvature}
\end{figure}

Finally, the curvature of the Bloch vector's trajectory is given by
\begin{equation}
\kappa = \left|\frac{d \hat{\bf t}}{ds}\right|
=\frac{|\ddot{\bf R}-\ddot{s}\,\hat{\bf t}|}{\left(\dot{s}\right)^2},
\label{eq:curvature}
\end{equation}
with the components $(A_X,A_Y,A_Z)$ of the acceleration
${\bf A}=\ddot{\bf R}$ given by
\begin{equation}
\left\{
\begin{array}{l}
{\displaystyle
A_X=-4Z\omega\cos(2\omega t),
}
\\
\\
{\displaystyle
A_Y=-4Z\omega\sin(2 \omega t),
}
\\
\\
{\displaystyle
A_Z=4\omega\,[X\cos(2\omega t)+Y\sin(2\omega t)],
}
\end{array}
\right.
\end{equation}
and 
\begin{equation}
\begin{array}{c}
{\displaystyle
\ddot{s}=4\,\left(\frac{d}{dt}\,|\cos(\omega t)|\right)
\sqrt{Z^2+[X\sin(\omega t)-Y\cos(\omega t)]^2}
}
\\
\\
{\displaystyle
+\frac{4|\cos(\omega t)|}{\sqrt{Z^2+[X\sin(\omega t)-Y\cos(\omega t)]^2}}\,
}
\\
\\
{\displaystyle
\times \frac{\omega}{2}\,[(X^2-Y^2)\sin(2\omega t)+XY\cos(2\omega t)].
}
\end{array}
\end{equation}
As shown in Fig.~\ref{fig:curvature}, the curvature diverges 
at the times $t_k$ of Eq.~(\ref{eq:tk}), corresponding to the cusps 
in the Bloch vector's path, i.e., when $\cos(\omega t)=0$.

\begin{thebibliography}{99}
\bibitem{qcbook_ex}
For a derivation of Hamiltonian 
(\ref{eq:hamc}) in the case of a single alkali atom 
in an electric field see, for instance, 
Ref.~\cite{qcbook},
exercise 8.5 page 532, solution page 632. 
The interaction of the valence electron with the electric 
field generated by a wave linearly polarized along the $z$-axis
reads, in the dipole approximation,
\begin{equation}
H_I(t)=-e z E(t),
\end{equation}
with
\begin{equation}
E(t)=E_0\cos(\omega t).
\end{equation}
The Rabi frequency is then given by
\begin{equation}
\Omega = \alpha D,
\end{equation}
where $\alpha=\frac{E_0}{2}$ and
$D=
-e\langle 0 |z |1\rangle$.
\bibitem{cauchypeano}
The fact that the functions $X(t)$, $Y(t)$ and $Z(t)$ are continuous
and differentiable functions is a consequence of the Cauchy-Peano
theorem applied to the first-order differential equations
(\ref{eq:noRWA1}).
\end{thebibliography}
\end{document}